\documentclass[prl, twocolumn, showpacs, superscriptaddress, amsmath,amssymb, floatfix, eqsecnum]{revtex4-1}
\pdfoutput=1
\usepackage{amsmath}
\usepackage{amssymb}
\usepackage{amsthm}
\usepackage{amsfonts}
\usepackage{bbm}
\usepackage{comment}
\usepackage[normalem]{ulem}
\usepackage{graphicx}
\usepackage[dvipsnames]{xcolor}
\usepackage{color,framed}
\usepackage{hyperref}
\usepackage{times}
\usepackage{enumerate}
\usepackage{lipsum}
\usepackage{slashed}
\usepackage{url}
\usepackage{bbm}
\usepackage{chngcntr}
\counterwithout{equation}{section}

\hypersetup{
    colorlinks=true, 
    linktoc=all,     
    linkcolor=blue,  
}

\def \beq {\begin{equation}}
\def \eeq {\end{equation}}
\def \beqa {\begin{eqnarray}}
\def \eeqa {\end{eqnarray}}
\def \bseq {\begin{subequations}}
\def \eseq {\end{subequations}}

\begin{document}

\title{Local diagnostics for strong-to-weak spontaneous symmetry breaking and non-equilibrium phase transitions}

\author{Carolyn Zhang}
\affiliation{Department of Physics, Harvard University, Cambridge, MA 02138, USA}

\begin{abstract}
We construct strongly $\mathbb{Z}_2$-symmetric local Markov/Lindblad dynamics exhibiting transitions between strong-paramagnetic behavior and strong-to-weak spontaneous symmetry breaking. In 1+1d, an absorbing-state construction gives a transition with scaling consistent with the parity-conserving branching-annihilating random-walk universality class. In 2+1d, a pair-flip variant of Toom’s rule provides evidence for a stable strong-paramagnetic/weakly symmetry-broken regime and a transition into an active SWSSB regime. We also introduce a marginal fidelity correlator of radius $R$, a local proxy for the usual SWSSB fidelity order parameter requiring tomography only on $\mathcal{O}(R^d)$-size regions. For broad classes of states, including symmetry-projected Gibbs-like states satisfying suitable local indistinguishability assumptions, we bound the error between the marginal and global fidelity correlators by terms decaying exponentially in $R$. These marginal fidelities provide a model-independent diagnostic of SWSSB in absorbing-state transitions, where microscopic notions of defects or activity are not universal.
\end{abstract}

\maketitle

\emph{\textbf{Introduction}}---There has been significant progress in the classification of mixed state phases from the perspective of \emph{states}: work in recent years has clarified how to determine when two states are ``roughly the same" and using these definitions, many different kinds of mixed state phases (including intrinsically mixed state phases) have been uncovered\cite{ma2023,sala2024,ma2025,ma2025doubled,lee2025,gu2025}. Typically, we say that two states $\rho_A$ and $\rho_B$ lie in the same phase if there is a low complexity operation, namely fast local Lindbladian evolution, to get from $\rho_A$ to $\rho_B$ and vice versa, possibly with additional requirements such as local reversibility and symmetry\cite{coser2019,ma2023,sang2025,sang2025mixed}. 
For the simplest case of strong $\mathbb{Z}_2$ symmetry, mixed states can demonstrate three phases: a strong paramagnetic (SP) phase, a strong-to-trivial spontaneous symmetry breaking (STSSB) phase, and an intrinsically mixed strong-to-weak spontaneous symmetry breaking (SWSSB) phase\cite{lessa2025} (see Ref.~\cite{supp} for a review). However, far less is known rigorously about the construction of parent Lindbladians, which are local, strongly symmetric Lindbladians with a particular density matrix as a steady state, with minimal steady state subspace given locality and symmetry constraints\cite{buvca2012,tibor2024,ziereis2025,shah2025,guo2025,liu2026}. Lindbladians may be viewed as open quantum system analogs of Hamiltonians.

One class of states for which there are well-known constructions of parent Lindbladians are Gibbs states. Gibbs samplers\cite{martinelli2004,kastoryano2016,chen2023,ding2025efficient} can indeed tune through thermal STSSB-SWSSB phase transitions if one restricts to symmetric jump operators. However, it is surprisingly hard to find an SP-SWSSB phase transition. An SP phase must have short-range fidelity correlator: $\lim_{|i-j|\to\infty }F_{ij}(\rho,O_iO_j^\dagger)=0$, since $F_{ij}(\rho,O_iO_j^\dagger)=\mathrm{Tr}(O_iO_j^\dagger\sqrt{\rho}O_i^\dagger O_j\sqrt{\rho})$ serves as an order parameter for the strong symmetry. Here, $O_i$ is a charged operator, so $O_iO_j^\dagger$ is neutral. On the other hand, any state of the form $(1+\prod_rX_r)e^{-\beta H}$ for a $\mathbb{Z}_2$ symmetric Hamiltonian $H$ and finite $\beta$ has nonzero fidelity correlator\cite{supp}. Therefore, to find SP-SWSSB phase transitions at finite tuning parameter, we would have to go beyond thermal states.

Previous work has indicated that in 1+1d, STSSB and SP phases are unstable to SWSSB: one can construct Lindbladians with steady states realizing the SP and STSSB phases, but it appears that generic strongly $\mathbb{Z}_2$ symmetric perturbations immediately take the steady states to an SWSSB phase\cite{shah2025}. One physical argument for this is as follows: suppose that we want to stabilize the SP fixed point state $\rho_0=|\{X_r=1\}\rangle\langle\{X_r=1\}|$. If we use strong $X$ decoherence to remove all states off diagonal in the $X$ basis, then the problem of stabilizing $\rho_0$ reduces to cleaning up $X_r=-1$ spins, which we will call errors. However, we cannot just flip all $X_r=-1$ spins; we have to annihilate them in pairs due to the strong $\mathbb{Z}_2$ symmetry. This is hard to do efficiently with local jump operators because the errors are point-like, so we typically get a diffusive timescale. On the other hand, generic strongly $\mathbb{Z}_2$ symmetric perturbations allow for the spontaneous creation of pairs of errors from the vacuum: $Z_rZ_{r+1}$ creates from a clean island of $X_r=1$ spins a pair of errors that can then diffuse through the system. The slow, diffusive cleaning of errors has little hope to overcome the immediate spawning of a finite density of errors. In the STSSB phase in higher dimensions, errors are domain walls, which can be efficiently cleaned up to obtain a stable finite temperature STSSB phase. However, the SP phase in any dimension has point-like errors and encounters the same instabilities.

\begin{figure}[t]
   \centering
   \includegraphics[width=.9\columnwidth]
   {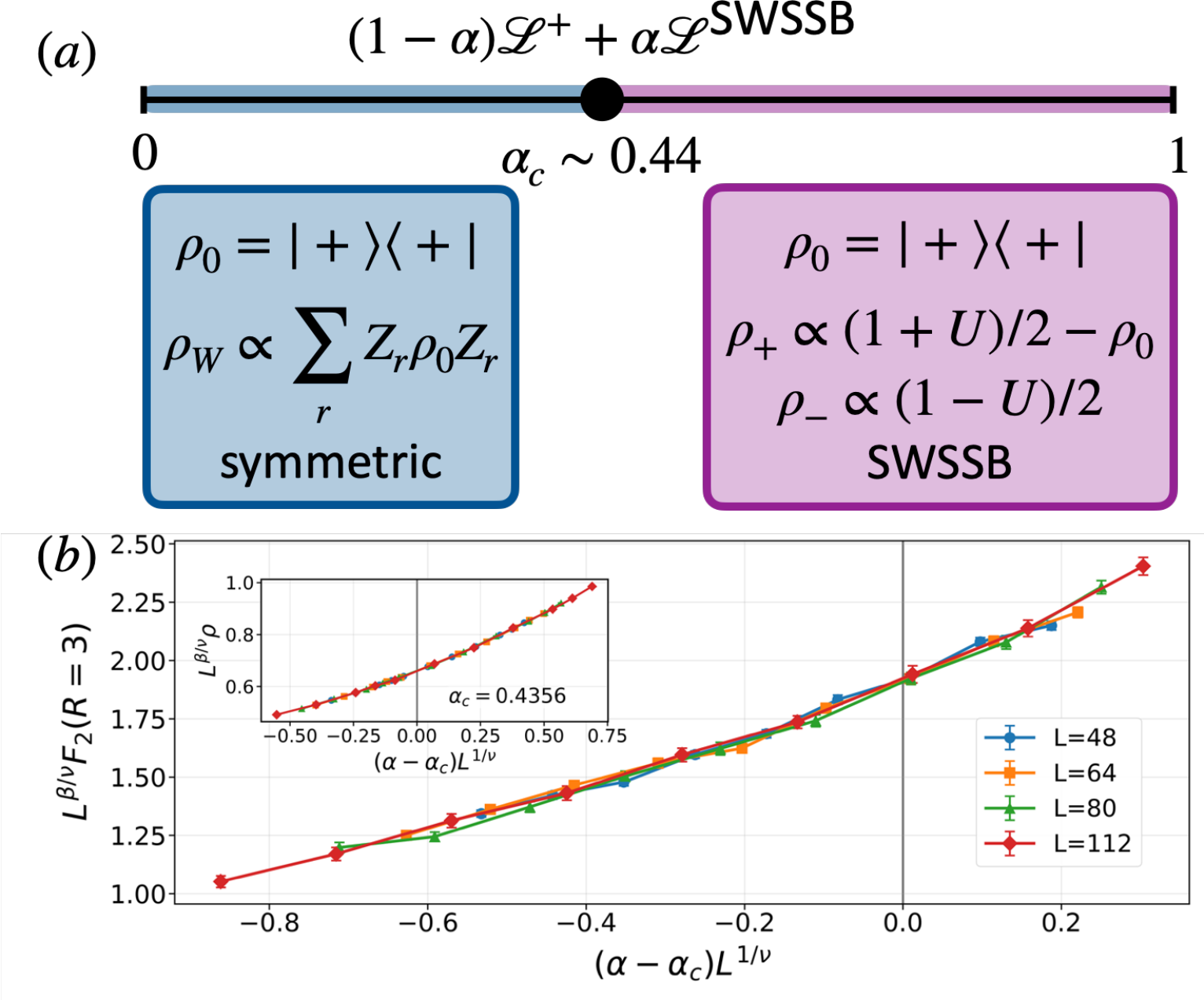}
   \caption{$(a)$ The 1+1d dynamics is designed to interpolate between $\mathcal{L}^+$ and $\mathcal{L}^{\text{SWSSB}}$. $\mathcal{L}^+$ has exact steady states $\rho_0$ and $\rho_W$ in the even and odd strong $\mathbb{Z}_2$ sectors respectively. $\mathcal{L}^{\text{SWSSB}}$ has an exact steady state space spanned by three steady states. All even sector states orthogonal to $\rho_0$ get taken to $\rho_+$ at late times, so $\rho_0$ is exponentially unlikely. $(b)$ The finite size scaling collapse gives critical exponents $\beta/\nu_{\perp}=0.53,1/\nu_{\perp}=0.56$ for the order parameter $n_-$ (inset) that match well with the PC-BARW universality class ($\beta/\nu_{\perp}=0.5, 1/\nu_{\perp}=0.54$ \cite{jensen1994}) with $\alpha_c=0.44$. The radius $R=3$ 2-point marginal fidelity correlator also demonstrates finite size scaling collapse with similar exponents: $\beta/\nu_{\perp}=0.52,1/\nu_{\perp}=0.57$, and $\alpha_c=0.46$. All data comes from the symmetry odd sector.} 
   \label{fig:fig1phasediag}
\end{figure}
In this work, we present examples of an SP to SWSSB phase transition for the $\mathbb{Z}_2$ symmetry $\prod_rX_r$ obtained by tuning parent Lindbladians. The dynamics is effectively classical, but we will say ``Lindbladian" to retain generality. The idea is to modify the SWSSB steady state space in a very modest way: in 1+1d, we add one additional steady state $\rho_0$ to the steady state space. Only $\rho_0$ goes to $\rho_0$ at late times; all states with no overlap with $\rho_0$ get taken to the SWSSB states $\rho_+\propto \frac{1+\prod_rX_r}{2}-\rho_0,\rho_-\propto \frac{1-\prod_rX_r}{2}$. Therefore, any random initial state would go to an SWSSB state with probability $1-2^{-L}$.  We will show that this slight modification of the steady state space will allow for a stable SP phase and a genuine SP to SWSSB transition in the familiar parity-conserving branching and annihilating random walk (PC-BARW) universality class\cite{grassberger1984,jensen1994,cardy1998,hinrichsen2000,odor2004}. In 2+1d, we end up having to add two more states to the steady state space: $\rho_0$ together with either $\rho_0^-=|\{X_r=-1\}\rangle\langle\{X_r=-1\}|$ or $\rho_{W}=\frac{1}{L^2}\sum_rZ_r\rho_0Z_r$ (we will later use $\rho_W$ to also refer to the analogous 1+1d version; 2+1d vs 1+1d should be clear from the context). We focus on the former case, which may give a new universality class of absorbing state transitions\cite{hinrichsen2000} based on a ``pair-flip Toom" update rule\cite{toom1980}. 

\emph{\textbf{SWSSB phase transition in 1+1d}}---First we present a very simple model of a phase transition in 1+1d between a strongly symmetric phase and an SWSSB phase of the $\mathbb{Z}_2$ symmetry $U=\prod_{r=0}^{L-1}X_r$, obtained by tuning a parameter in a Lindbladian. We will turn on strong $X$ decoherence that quickly makes any input state diagonal in the $X$ basis, so we only need to specify classical dynamics in the $X$-basis. In this basis, we have two classical Markov processes. First, we have the absorbing rule 
\begin{align}
\begin{split}
    \mathcal{L}_r^+:&(++)_{r-1,r},(---)_{r-1,r}\to (++)_{r-1,r}\\ &(+-)_{r-1,r},(-+)_{r-1,r}\to (-+)_{r-1,r}
\end{split}
\end{align}
Intuitively, this update rule leaves states with $X_r=+1$ alone and pushes $X_r=-1$ errors to the left. Our SWSSB rule is (with $r$ being the middle site)
\begin{align}
\begin{split}
   \mathcal{L}_r^{\text{SWSSB}}: &(+++)\to (+++)\\ &\text{ else scramble in sector}\setminus (+++)
\end{split}
\end{align}
meaning we exclude the configuration $+++$ but otherwise we include all transitions between configurations with the same $\prod X=\pm 1$ on three sites. In particular, we include transitions between configurations with one $-$ spin and $(---)$. We can then consider the interpolating Lindbladian
\begin{equation}
\mathcal{L}(\alpha)=(1-\alpha)\mathcal{L}^++\alpha\mathcal{L}^{\text{SWSSB}}
\end{equation}

At $\alpha=0$, we get dynamics that stabilizes one steady state in each of the two symmetry sectors. In the $U=+1$ sector, the steady state is $\rho_0$ while in the $U=-1$ sector, the steady state is $\rho_W$. Even though $\rho_W$ has long range mutual information, and is therefore not two-way local channel connected to a product state, we will say that this Lindbladian realizes an SP phase because the $U=+1$ steady state is clearly SP. It is also not hard to check that the steady state subspace at $\alpha=1$ is generated by $\rho_0,\rho_+,$ and $\rho_-$ with all states except for $\rho_0$ going to the SWSSB states $\rho_{\pm}$ as $t\to\infty$; we show this rigorously in Ref.~\cite{supp} by checking ergodicity and double stochasticity within the even sector without $\rho_0$ and in the odd sector. While giving only a mild modification at very late times, the addition of $\rho_0$ to the steady state subspace dramatically affects the transient dynamics of $\mathcal{L}^{\text{SWSSB}}$: it makes the mixing time $\mathcal{O}(L)$ because errors cannot spawn from a clean state, so an initial state with single error surrounded by spins in the $+$ state would take $\mathcal{O}(L)$ time to get close to $\rho_-$. 

$\mathcal{L}(\alpha)$ includes $2A\to\emptyset$, $A\to 3A$ transitions, where $A$ is an error, so it is not surprising that it demonstrates a critical point in the PC-BARW universality class. The typical order parameter in the literature for this transition is the ``defect density" $n_-=\frac{1}{L}\sum_r\frac{1-\langle X_r\rangle }{2}$ (we will also call this error density). We show in Fig.~\ref{fig:fig1phasediag} that $\mathcal{L}(\alpha)$ demonstrates a transition at $\alpha\sim 0.44$ with critical exponents consistent with those in the PC-BARW class.

Note that the Lindbladian is $\mathcal{O}(1/L^2)$ gapless at $\alpha=0$ and likely remains so for the entire range of $\alpha\in[0,\alpha_c)$. For $\alpha>\alpha_c$, the model is likely gapped above its exponentially long-lived SWSSB steady states. At $\alpha=1$, these metastable SWSSB states become exact steady states.
\begin{figure}[t]
   \centering
   \includegraphics[width=.9\columnwidth]
   {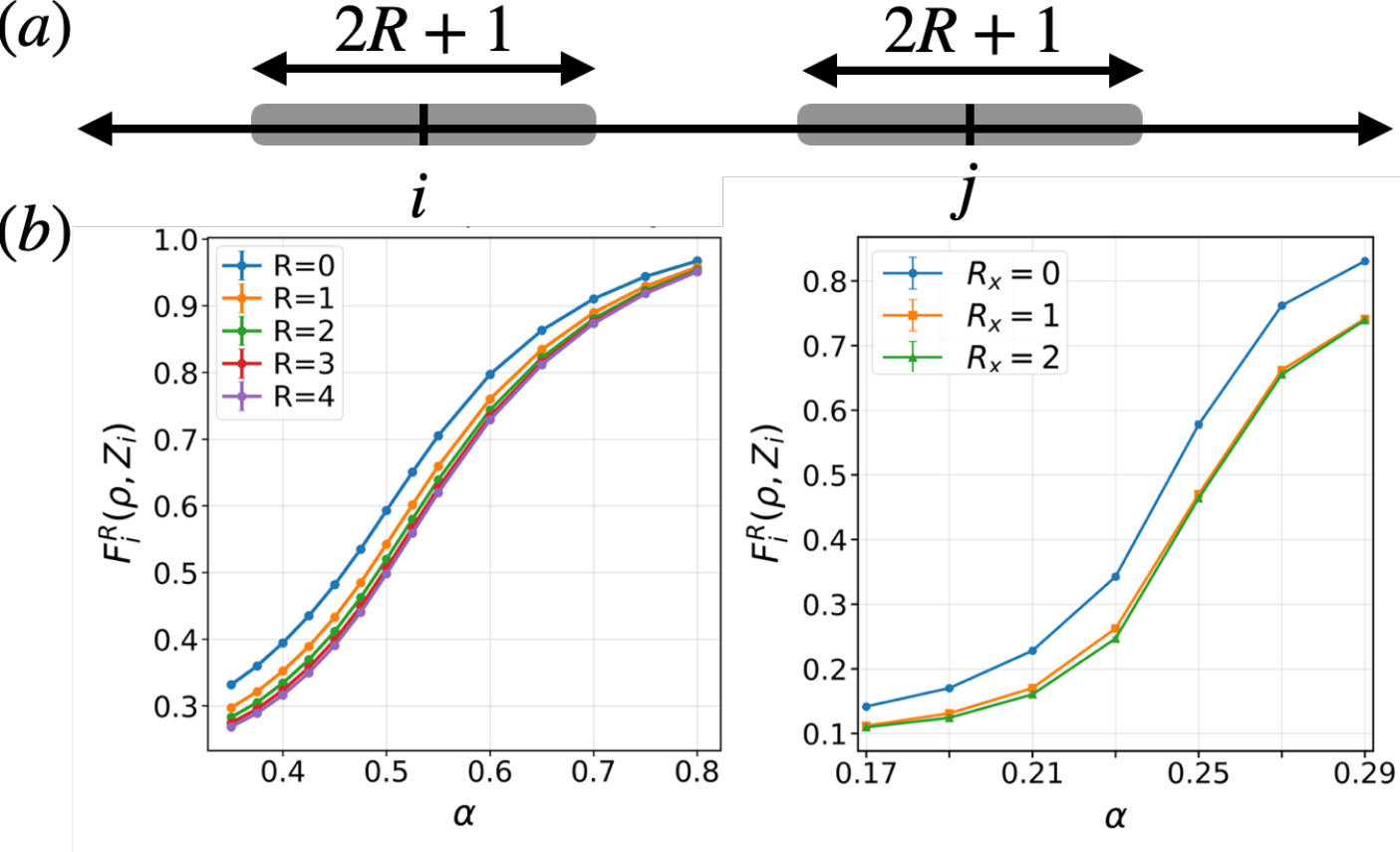}
   \caption{$(a)$ The marginal one-point fidelity correlator takes the reduced density matrix on $R(i)\cup R(j)$, which includes $2(2R+1)$ sites. In higher dimensions, we specify $R_x$ and $R_y$, and take the reduced density matrix on a region of size $(2R_x+1)(2R_y+1)$ around $i$ and $j$. $(b)$ The fidelity correlator as a function of $R$ and $R_x$ shows rapid convergence; see Ref.~\cite{supp} for calculations supporting faster convergence in the SWSSB phase and in the 2+1d case for $R_x>0$ away from $\beta_c$. The first plot tracks radius $R$ fidelity correlator for a system of size $L=112$ across the 1+1d phase transition. The second plot tracks $R_x$ fidelity correlator for a system of size $L^2=400$ across the 2+1d strict pair flip Toom transition, with $R_y=1$ fixed (i.e. for $R_x=0,1,2$ correspond to patches with $3,9,15$ sites around $i$ respectively). Because the system is translation invariant, we also perform additional averaging over choices of $i$, together with averaging over 50-100 runs and 200-400 late time samples per run. } 
   \label{fig:fig2fidelity}
\end{figure}

\emph{\textbf{Marginal fidelity}}---To diagnose SWSSB, one typically uses a fidelity correlator\cite{lessa2025}. In this work, we will use the Holevo fidelity $F_{ij}(\rho,O_iO_j^\dagger)$ defined in the Introduction primarily with $O_i=Z_i$ a local operator charged under the strong $\mathbb{Z}_2$ symmetry\cite{kholevo1972,wilde2018,weinstein2025,liu2025}. In the SWSSB phase, $\lim_{|i-j|\to\infty}F_{ij}(\rho,Z_iZ_j)\neq 0$ (more generally, we just need $F_{ij}(\rho,O_iO_j^\dagger)$ to be nonzero for some quasi-local operators $O_i,O_j^\dagger$ charged under the symmetry).

In classical systems, the fidelity reduces to a Bhattacharyya distance between probability distributions: $F_{ij}(\rho,Z_iZ_j) = \sum_{\{X_r\}}\sqrt{p(\{X_r\})p(\{X_r^{ij}\})}$ where $p(\{X_r^{ij}\})$ is the configuration $\{X_r\}$ with spins $i$ and $j$ flipped. Therefore, if we collect several snapshots of the system at late times, we just need to ask: given a snapshot $\{X_r\}$, is there also a snapshot $\{X_r^{ij}\}$? There is, however, a sampling issue that makes it easy to underestimate $F_{ij}(\rho,Z_iZ_j)$: collecting $N$ snapshots would make it very unlikely to see $\{X_r^{ij}\}$ even if $p(\{X_r^{ij}\})>0$, if $N/2^{V}<1$ (where $V=L^d$ is the volume in $d$ spatial dimensions). To alleviate this sampling issue, we introduce in this section a \emph{marginal fidelity} with radius $R$, defined as
\begin{equation}
    F_{ij}^R(\rho,Z_iZ_j)=\mathrm{Tr}(\sqrt{\rho_{R(i)\cup R(j)}}Z_iZ_j\sqrt{\rho_{R(i)\cup R(j)}}Z_iZ_j)
\end{equation}
where $\rho_{R(i)\cup R(j)}=\mathrm{Tr}_{\overline{R(i)\cup R(j)}}(\rho)$ is the reduced density matrix on the $2(2R+1)^d$ sites $R(i)\cup R(j)$ (see Fig.~\ref{fig:fig2fidelity}.$(a)$). The 2-point marginal fidelity only requires $\mathcal{O}(2^{2(2R+1)^d})$ snapshots, and more generally in quantum systems it requires tomography on $2(2R+1)^d$ sites. In Ref.~\cite{supp} we also show that
\begin{equation}
    \left|F_{ij}^R(\rho,Z_iZ_j)-F_i^R(\rho,Z_i)F_j^R(\rho,Z_j)\right|\leq2(2I(A:C)_{\rho})^{1/4}
\end{equation}
where $F_i^R(\rho,Z_i)$ is the marginal one-point correlator with radius-$R$ and $I(A:C)$ is the mutual information with $A=R(i),C=R(j)$. If $I(A:C)$ is exponentially decaying in the distance between $A$ and $C$, we can get away with only $\mathcal{O}(2^{(2R+1)^d})$ classical snapshots, or tomography on $(2R+1)^d$ sites. Note that factorization of $F_{ij}^R(\rho,Z_iZ_j)$ often occurs even when $\rho$ does not demonstrate exponential decay of $I(A:C)$\cite{supp}. 

Why should we expect that $F_{ij}^R(\rho,Z_iZ_j)$ serves as a good proxy for SWSSB? Intuitively, $F_{ij}(\rho,Z_iZ_j)$ indicates whether or not there are local fluctuations in $X_i,X_j$: fixing $\{X_r|r\neq i,j\}$, is there nonzero probability for both $X_i,X_j$ and $-X_i,-X_j$? The marginal fidelity correlator similarly indicates local fluctuations, albeit conditioning on a subset of sites surrounding $i,j$ rather than the entire complement of $i,j$. To make this intuition more precise, we can derive several properties of $F_{ij}^R(\sigma,Z_iZ_j)$ using the fact that it satisfies a quantum data processing inequality 
\begin{equation}
\mathrm{Tr}\left(\sqrt{\mathcal{E}(\rho)}\sqrt{\mathcal{E}(\sigma)}\right)\geq \mathrm{Tr}\left(\sqrt{\rho}\sqrt{\sigma}.\right)
\end{equation}
where $\sigma=Z_iZ_j\rho Z_iZ_j$, for any CPTP map $\mathcal{E}$. In particular, choosing $\mathcal{E}$ to be a partial trace over $\overline{R(i)\cup R(j)}$ immediately gives the following: 
\begin{itemize}
\item \emph{Monotonicity: }$F_{ij}^R(\rho,O_iO_j^\dagger)\leq F_{ij}^{R'}(\rho,O_iO_j^\dagger)$ if $R>R'$. In particular, if the state has SWSSB so $F_{ij}(\rho,O_iO_j^\dagger)>0$, then $F_{ij}^R(\rho,O_iO_j^\dagger)>0$ as long as $R(i)\cup R(j)$ includes the full support of $O_iO_j^\dagger$ (i.e. $R\geq 0$ for $O_iO_j^\dagger=Z_iZ_j$). In addition, if $F_{ij}^R(\rho,O_iO_j^\dagger)=0$ for some $R\leq \mathcal{O}(L)$, then the state certainly does not have SWSSB.
\item \emph{Stability: }If $F_{ij}^R(\rho,O_iO_j^\dagger)\sim\mathcal{O}(1)$ for some $O_i$ supported on $n$ sites, then $F_{ij}^R(\mathcal{E}[\rho],\tilde{O}_i\tilde{O}_j^\dagger)>0$ for some dressed charged operator $\tilde{O}_i$ supported on $n+d$ sites where $d$ is the operator spreading length proportional to the depth of the symmetric local channel $\mathcal{E}$\cite{lessa2025} as long as $n+d\leq 2R+1$. Assuming that $\tilde{O}_i$ has some overlap with $O_i$, we would also expect that $F_{ij}^{R}(\mathcal{E}[\rho],O_iO_j^\dagger)>0$. This follows from the stability theorem of Ref.~\cite{lessa2025} for the Uhlmann fidelity $\tilde{F}_{i,j}(\rho,\sigma)$ together with the fact that $\tilde{F}_{i,j}(\rho,Z_iZ_j)^2\leq F_{ij}(\rho,Z_iZ_j)\leq \tilde{F}_{i,j}(\rho,Z_iZ_j)$.
\end{itemize}
In Ref.~\cite{supp}, we also show that the marginal fidelity, similar to the true fidelity, implies long-range conditional mutual information (CMI):
\begin{itemize}
\item \emph{Complexity: } If $\rho$ is strongly symmetric then $F_{i}^R(\rho,O_i)$ lower bounds the CMI $I(A:C|B)$: $F_i^R(\rho,O_i)\leq 2\sqrt{2}I(A:C|B)^{1/4}$. Therefore if $F_i^R(\rho,O_i)>0$ uniformly even as $R\sim\text{poly}\log L$, the state has long range CMI.
\end{itemize}

Finally, we show in the supplemental material~\cite{supp} a general lower bound on the fidelity for (symmetry projected) quantum Gibbs states: 
\begin{equation}
    F_{ij}(\rho,Z_iZ_j)>e^{-\beta \|V_i+V_j\|/2}
\end{equation}
where $V_i+V_j=Z_iZ_jHZ_iZ_j-H$ are local operators with bounded norm. We also show that any state with exponential decay of CMI gives symmetry projected states satisfying
\begin{equation}
    |F_{ij}^R(\rho,Z_iZ_j)-F_{ij}(\rho,Z_iZ_j)|\leq \mathcal{O}(e^{-R/\xi})
\end{equation}
under certain local indistinguishability conditions.

\emph{\textbf{Marginal fidelity in the 1+1d phase transition}}---The conventional order parameter for the 1+1d PC-BARW phase transition is the error density $n_-$. Here we also investigate the behavior of the marginal two-point fidelity. Numerically, we find an excellent agreement between the two-point fidelity and the $n_-$ order parameter in terms of critical exponents and $\alpha_c$ (see Fig.~\ref{fig:fig1phasediag}). In the absorbing phase in the odd sector, $n_-\sim\frac{1}{L}$ and we find by explicit calculation\cite{supp} that $F_{ij}^R(\rho,Z_iZ_j)\sim\frac{1}{L}$\cite{supp} without much $R$ dependence. In the active phase, $F_{ij}^0(\rho,Z_iZ_j)\sim 4n_-(n_--1)\sim 2n_-$ for $n_-\ll 1$, if we assume that the sites are uncorrelated. This supports the case for $F_{ij}^R(\rho,Z_iZ_j)$ behaving very similarly to $n_-$. On the other hand, the one-point correlator tracks $n_-$ more poorly in the SP phase and therefore gives critical exponents that differ from those of $n_-$.


In the symmetry even sector, at timescales exponentially large in $L$, we expect that all states eventually fall into the absorbing state $\rho_0$ for $\alpha_c<\alpha<1$. We can therefore take as an ansatz $\rho(s)=(1-s)\rho_{\alpha}+s\rho_0$ where $s$ is the survival probability. In Ref.~\cite{supp} we show that in this case, $F_{ij}^R(\rho,Z_iZ_j)$ approximates $F_{ij}(\rho,\sigma)$ up to error exponentially small in $R$ as long as $s\gg\frac{1}{1+2^{4R}}$.



\emph{\textbf{2+1d phase transitions}}---The naive generalization of the 1+1d update rules to 2+1d does not give a stable absorbing phase, according to the renormalization group analysis of Ref.~\cite{cardy1998}. 
Adding additional particle-based absorbing dynamics should not change this result because processes like $3A\to A$, $4A\to A$ are higher order processes compared to $2A\to\emptyset$: they require $3$ or $4$ errors to meet together simultaneously.

To get a genuine SP phase, we have two options: one option is to weaken the SWSSB dynamics by only allowing $2A\to 4A$ branching but not $A\to3A$ branching. 
This puts branching and annihilation on the same footing: both are controlled by the probability that two errors meet together. 
Disallowing $1\to 3$ branching forces the SWSSB side to have at least one other non-SWSSB steady state in addition to $\rho_0$. Because the single error sector $n_-=1/L^2$ is preserved under the dynamics, $\rho_W$ is forced to be an exact steady state. However, a random $X$ basis product state still goes to the SWSSB states with probability $1-(1+L^2)/2^L\sim1-\mathcal{O}(e^{-L})$. Interpolating between SP dynamics and pair activated SWSSB dynamics typically gives a first order nucleation-based transition, because it requires a sufficiently dense bubble of errors to initiate activity\cite{binder1987,hinrichsen2000}.

\begin{figure}[t]
   \centering
   \includegraphics[width=.9\columnwidth]
   {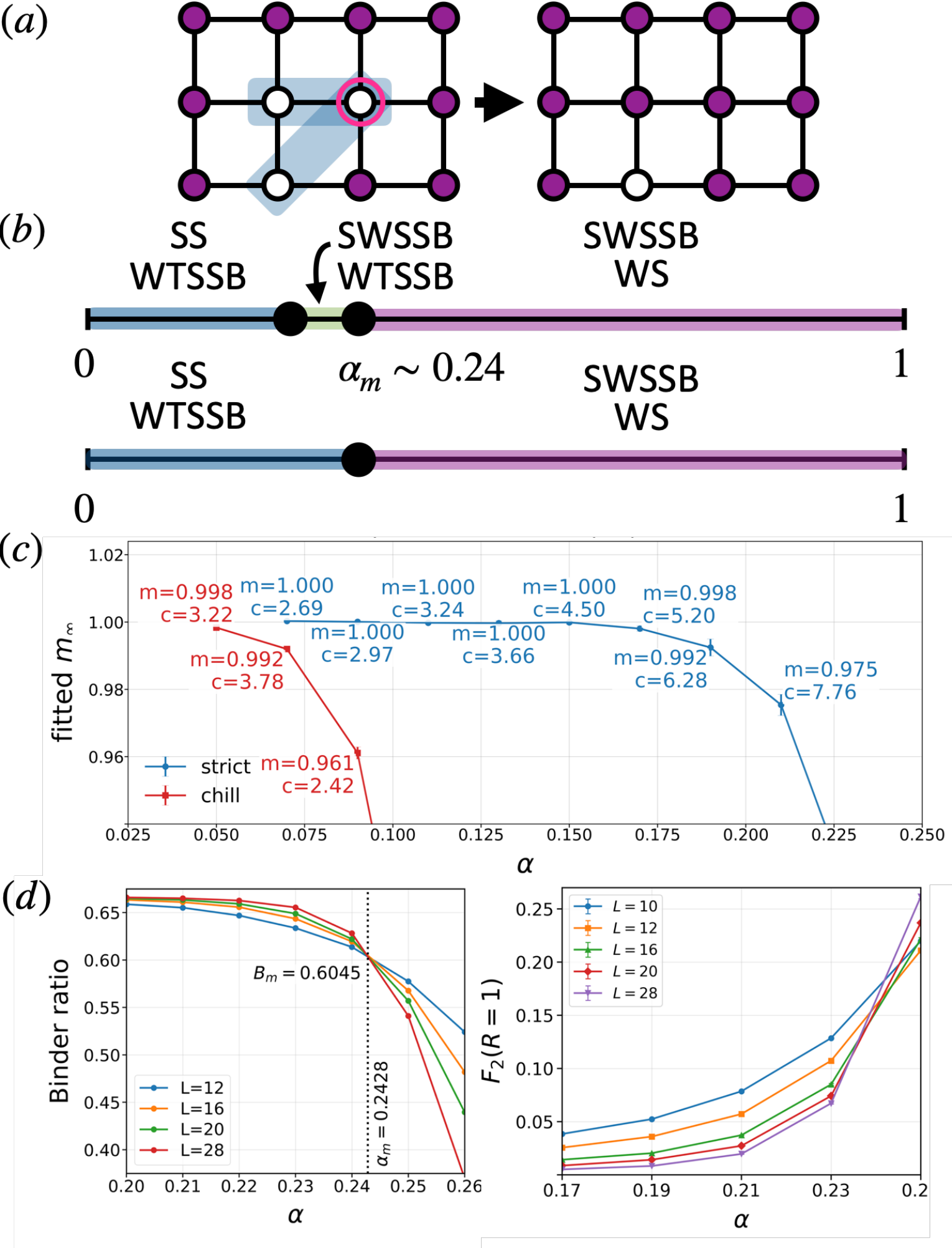}
   \caption{$(a)$ The purple circles represent $+$ spins and white circles represent $-$ spins, and the blue links represent pair flips allowed for strict pair-flip Toom. $(b)$ Two possible phase diagrams for the 2+1d model with strict pair-flip Toom. $(c)$ In the odd sector in the SP phase, we should have $\langle|m|\rangle=m_{\infty}-c/L^2$ with $m_{\infty}=1$ and $c/2$ giving the size of a wandering $\mathcal{O}(1)$ minority domain, so we fit $m_{\infty}$ and $c$ using $L=10,12,16,20$ data (in the plot we use $m$ instead of $m_{\infty}$ to save space). The strict pair flip Toom rule (blue) demonstrates a convincing stable phase where $m_{\infty}=1$ whereas the chill pair flip Toom rule has $m_{\infty}$ very close to but not quite 1 for very small $\alpha$. We also obtain data points (not shown in the plot for clarity) $(\alpha=0.01,m_{\infty}=1.000)$, $(\alpha=0.02,m_{\infty}=0.9999)$, $(\alpha=0.03,m_{\infty}=0.9993)$, and $(\alpha=0.04,m_{\infty}=0.999)$ for chill pair flip Toom. $(d)$ The plot of the Binder ratio $1-\frac{\langle m^4\rangle}{3\langle m^2\rangle^2}$ is $2/3$ as expected for small $\alpha$, and demonstrates a clean crossing at the WTSSB to WP transition at $\alpha_m=0.2428$. The $R=1$ two-point fidelity correlator demonstrates the expected behavior of decreasing with $L$ at small $\alpha$ and increasing with $L$ at large $\alpha$, but the critical point appears to be too strongly affected by finite size effects to get a convincing $\alpha_c$ and set of critical exponents. Other possible order parameters similar to error density behave in the same way\cite{supp}.} 
   \label{fig:fig32d}
\end{figure}
Instead of weakening the SWSSB dynamics, we will focus on strengthening the SP dynamics by making the SP dynamics domain-wall based. The SWSSB dynamics must preserve both $\rho_0$ and $\rho_0^-$, so we choose dynamics in the $X$ basis that picks a plaquette $p$ and updates
\begin{align}
\begin{split}
\mathcal{L}^{\text{SWSSB}}_p:&(++++)\to (++++)\\
&(----)\to (----)\\
&\text{else scramble in sector}\setminus \{(++++), (----)\}
\end{split}
\end{align}
meaning we exclude the $(++++),(----)$ configurations but otherwise include all transitions between configurations of the same $\prod_{r\in p}X_r$. In particular, the eight configurations with two $-$ spins get scrambled between each other but are not mixed with $(++++),(----)$. For the SP dynamics, we need a Lindbladian that stabilizes both $\rho_0$ and $\rho_0^-$ and erodes domain walls in a strongly symmetric way. We use an update rule based on Toom's rule, which is known to stabilize a classical memory in 2+1d against both symmetric and symmetry breaking noise\cite{toom1980}. The usual Toom's rule is as follows: pick a site and if it antialigns with both its north and east neighbors (``Toom unstable"), flip it. To preserve the strong symmetry, we need to flip spins in pairs, so we implement two versions of a pair-flip Toom's rule. If a site is Toom unstable, then we compute the change in domain wall length $\Delta n_{\mathrm{dw}}$ associated with flipping the spin together with one of its eight neighbors and choose a partner among those that give $\Delta n_{\mathrm{dw}}<0$ (``strict" pair flip Toom) or $\Delta n_{\mathrm{dw}}\leq 0$ (``chill" pair flip Toom) (see Fig.~\ref{fig:fig32d}.$(a)$ for an example and \cite{supp} for more details). The change in domain wall length can be computed locally around the site, so this rule can be implemented by a local Lindbladian. We find numerical evidence that the strict pair-flip Toom rule does appear to have a stable SP phase, while the chill one either does not or has a very small $\alpha_c<0.02$ (see Fig.~\ref{fig:fig32d}.(c)). The intuition behind this is as follows: pair-flip Toom tries to accumulate aligned spins into large domains, which it can then aggressively erode. The objects dangerous to the erosion are small islands that wander off on their own and cannot be annihilated due to the strong symmetry. We give evidence in Ref.~\cite{supp} that compared to the strict pair-flip Toom rule, the chill pair-flip Toom rule allows much more peeling off and wandering of small islands. While in the $+1$ sector the steady states look like those of Toom's rule in the $X$ basis, the strong symmetry gives contraints on the dynamics and forces additional steady states in the symmetry odd sector.

Note that strict pair-flip Toom on its own has many other steady states in addition to the desired $\rho_0$ and $\rho_0^-$ states: any isolated minority spin and any rectangular domain with side length $\geq 3$ is also completely frozen. Therefore, the strict pair-flip Toom rule on its own is not a good parent Lindbladian for an SP phase. However, adding a small amount of SWSSB dynamics removes the frozen states: it is not hard to see that the SWSSB transitions allow for all of the isolated minority spin states to be connected with each other. The numerics in Fig.~\ref{fig:fig32d}.$(c)$ show that the sample-averaged absolute value of magnetization $\overline{\langle |m|\rangle}$ converges to precisely 1 for $\alpha\leq 0.15$, so the SWSSB dynamics remove \emph{all} of the spurious steady states of strict pair-flip Toom. The convergence to precisely 1 also indicates that there cannot be fluctuations at late times for those values of $\alpha$, meaning there is no SWSSB and the system realizes an SP phase at least for $\alpha\leq 0.15$.

Due to the domain-wall based dynamics, which preserves a $X_r\to -X_r$ symmetry, the above Lindbladian comes with weak $\mathbb{Z}_2$ $\prod_rZ_r$ symmetry. While the pair-flip Toom rule stabilizes a WTSSB phase, the SWSSB rule gives a weak paramagnetic (WP) phase. In Fig.~\ref{fig:fig32d}.$(d)$, we show that the Binder cumulant plot demonstrates a clean crossing indicating a WTSSB to WP phase at $\alpha_m\sim 0.24$. 
We measure various quantities similar to error density: $\rho_{\mathrm{act}}$ is the number of plaquettes not in the $(++++)$ or $(----)$ configuration and $n_{\mathrm{dw}}$ is the domain wall length (see \cite{supp}). However, there is too much finite size error to give a convincing determination of $\alpha_c$, so we leave open the possibility of a critical region, a direct ``deconfined critical point"-like transition between SP/WTSSB and SWSSB/WP, or an intervening SWSSB/WTSSB phase (see Fig.~\ref{fig:fig32d}.$(b)$). In fact, this splitting of transitions with an intervening SWSSB/WTSSB phase is generic in a strongly symmetric version of Glauber dynamics, which measures $X_rX_{r'}$ and flips spins in pairs. Zero temperature Glauber dynamics has the same $\mathrm{span}(\rho_0,\rho_0^-)$ steady state space in the symmetry even sector. However, at any nonzero $T$ the strong symmetry immediately becomes SWSSB. The WTSSB to WP phase transition occurs later, at the $T_c$ of the 2d classical Ising model. 
A phase with SP phase coexisting with a WP phase may not be possible without spontaneously breaking the weak translation symmetry due to the mixed strong-weak Lieb-Schultz-Mattis anomaly\cite{lessamultipartite}. 

\emph{\textbf{Discussion}}---The phase transitions in 1+1d and 2+1d discussed in this work can easily be adapted to give phase transitions between other phases and SWSSB. For example, performing Kramers-Wannier duality on all of the local jump operators (which maintains locality due to the strong symmetry) would give a Lindbladian tuning a STSSB to SWSSB phase transition in 1+1d. If we in addition replace the bare $X_r$ with the patch operator of the anomalous $\mathbb{Z}_2$ symmetry in 1+1d\cite{chatterjee2023,zhang2023}, we get a phase transition between STSSB and SWSSB of the anomalous $\mathbb{Z}_2$ symmetry\cite{lessamultipartite}. In 2+1d, performing the same gauging map on the jump operators for the strict pair flip Toom rule appears to give a stable phase absorbing to Toric code states with $\{A_v=+1\}$ and $\{A_v=-1\}$, stable to noise that can only proliferate anyon pairs from existing anyon errors. 

In 2+1d we have an ambiguous phase diagram (see Fig.~\ref{fig:fig32d}) that asks for more in-depth numerical exploration. It would be particularly interesting if one can prove rigorously the stability of the SP phase with certain classes of dynamics against noise that preserves the absorbing state (so cannot nucleate errors for a clean state). It would also be interesting to see if chill pair-flip Toom is really unstable to any $\mathcal{O}(1)$ rate of $A\to3A$ branching unlike strict pair-flip Toom, and if so to understand the physics of the instability. More generally one can also tune between these two limits, allowing some probability of accepting $\Delta n_{\mathrm{dw}}=0$ moves. In the equilibrium context, we are familiar with stabilizing phases by imposing symmetries. In this case the ``symmetry" appears to include requiring jump operators to commute with the absorbing state $\rho_0$, in particular disallowing $\emptyset\to 2A$ error nucleation. As noted in the 2+1d discussion, forbidding $A\to 3A$ branching can also be thought of as, perhaps more universally, as adding an extra steady state $\rho_W$, and our pair-flip Toom approach also involved adding an additional steady state $\rho_0^-$. These perspectives may be useful for constructing more general phase diagrams and stability-preserving perturbations. 

In contrast to 2+1d, in 1+1d it may be impossible to get an intervening SWSSB with WTSSB phase where the strong and weak symmetries have a mixed anomaly with the weak translation symmetry. In 1+1d, the canonical example of such a phase $(1+\prod_rX_r)e^{-\beta H}$ for $\beta>\beta_c$ would not work because there are no finite temperature SSB phases. If this phase does not exist then there may either be no stable SP phase or an enforced ``DQCP". 

Another approach to stabilizing an SP phase is to use a parent Lindbladian with $\mathrm{poly}\log L$ local jump operators. Previous work has shown that the SP phase is stable against weak symmetric decoherence, meaning it can be recovered by a quasilocal Petz recovery map\cite{sang2024,sang2025,sang2025mixed}. However, this recovery generally involves dynamics with $\mathrm{poly}\log L$ jump operators. It is likely that having a Lindladian with a small amount of generic symmetric noise together with the recovery Lindbladian would stabilize a state that is in the SP phase but not necessarily at the SP fixed point. Presumably there will be a threshold where the noise overcomes the recovery, leading to SWSSB phase, but the phase transition may not be clean due to the $\mathrm{poly}\log L$ locality of the jump operators. 

There are many interesting possible phases and phase transitions to explore in open quantum many-body systems, especially when we consider having symmetries and multiple absorbing states that may have quantum coherences\cite{marcuzzi2016,ha2025,wampler2025}. In particular, it would be interesting to understand the applicability of the (marginal) fidelity correlators in this context: how do we compare them to more traditional order parameters, and understand their critical behavior? Even though the classification of mixed states has seen a lot of progress, there is more to be understood in the context of parent Lindbladians: how are we allowed to construct steady state spaces of local Lindbladians? What kind of constraints arise from locality and symmetry? Can we have different behavior in different symmetry sectors? To explore these questions, the (marginal) fidelity correlator may be an important tool due to its information theoretic properties and its universal applicability. 

\emph{Acknowledgements}---We thank Sarang Gopalakrishnan, Alexey Khudorozhkov, and Zack Weinstein for helpful discussions. We acknowledge the use of ChatGPT 5.5 in the writing of code and various derivations in the supplemental material. C.Z. is supported by the Harvard Society of Fellows. 

\emph{Note added}---We thank the authors of Refs.~\onlinecite{chong,else} for coordinating submission of papers that will appear in the same arXiv posting.

\bibliography{lindbladian}
\onecolumngrid        




\end{document}


\title{Supplemental Material: \\ Local diagnostics for strong-to-weak spontaneous symmetry breaking and non-equilibrium phase transitions}

\author{Carolyn Zhang}
\affiliation{Department of Physics, Harvard University, Cambridge, MA 02138, USA}

\maketitle
\tableofcontents

\section{Review of strong/weak symmetries and phase classification}\label{sreview}
In this section we review mixed state symmetries and some key examples of strongly symmetric density matrices for the different phases discussed in the main text. A mixed state is said to have a strong unitary symmetry if 
\begin{equation}
    U\rho=e^{i\theta}\rho\qquad \rho U^\dagger = e^{-i\theta} \rho
\end{equation}
and a state has a weak symmetry if 
\begin{equation}
    U\rho U^\dagger = \rho
\end{equation}
In the weak symmetry case, $\rho$ may consists of a mixture of states in different symmetry sectors of $U$. A Lindbladian is a superoperator of the form
\begin{equation}
    \mathcal{L}[\rho]=-i[H,\rho]+\sum_jL_j\rho L_j^\dagger-\frac{1}{2}\{L_j^\dagger L_j,\rho\}
\end{equation}
and generates time evolution: $\dot{\rho}=\mathcal{L}[\rho]$. The form above guarantees that it generates a completely positive trace-preserving map. We will consider \emph{local} Lindbladians, meaning the jump operators $L_j$ are local (in fact, they are finite range in all of the examples in this paper). We will not need the Hamiltonian $H$ in this work; the dynamics we describe are purely dissipative. 

We say that a Lindbladian has a strong unitary symmetry $U$ if
\begin{equation}
    [H,U]=[L_j,U]=0\forall j
\end{equation}

A sufficient condition for weak symmetry is that 
\begin{equation}
    \text{either}\qquad UL_jU^\dagger = L_{j'}\qquad\text{or}\qquad UL_jU^\dagger = e^{i\theta }L_j
\end{equation}
for each $j$. Every strong symmetry is also a weak symmetry, but a weak symmetry may not be a strong symmetry.

If a Lindbladian has a strong symmetry, then it must have at least one steady state in each symmetry sector. In particular, for $\mathbb{Z}_2$ symmetry, there must be a steady state in the $+1$ sector and the $-1$ sector. This is because the dynamics preserves the symmetry sector, and there are initial states that can be in either symmetry sector. For example in the main text for the 1+1d example, the SP dynamics stabilized $\rho_0=|\{X_r=1\}\rangle\langle\{X_r=1\}|$ in the $+1$ sector and $\rho_W=\frac{1}{L}\sum_rZ_r\rho_0Z_r$ in the $-1$ sector. While the state in the $+1$ sector is a trivial product state, the state in the $-1$ sector has long range mutual information/classical correlations.

In 1+1d, we have key fixed point states for the three phases:
\begin{equation}
    |\{X_r=1\}\rangle\langle\{X_r=1\}|\qquad |\mathrm{GHZ}\rangle\langle\mathrm{GHZ}|\qquad \frac{1+U}{2^L}
\end{equation}
The first state is a representative of the strong paramagnetic (SP) phase. The second state is a representative of the strong-to-trivial SSB (STSSB) phase, and the last state is a representative of the strong-to-weak SSB (SWSSB) phase. While the first two states can be evolved into $\frac{1+U}{2^L}$ in $\mathcal{O}(\mathrm{poly}\log L)$ time with a strongly symmetric local Lindbladian, the reverse process cannot occur on timescales $<\mathcal{O}(L)$, so the three states lie in distinct phases. It is easy to see that the second and third state have long range fidelity correlator: $F_{ij}(\rho,Z_iZ_j)=1$ regardless of $|i-j|$. The fidelity correlator diagnoses strong symmetry breaking, but does not imply weak symmetry breaking. 

In 2+1d, we also mentioned the state
\begin{equation}
    (1+U)e^{-\beta H}/Z
\end{equation}
where $H$ commutes with $U$ as well as a weak $\mathbb{Z}_2$ $\prod_rZ_r$ symmetry. This state can demonstrate long range order, since $e^{-\beta H}/Z$ can describe a weak-to-trivial (WTSSB) phase if i.e. $H=-\sum_{r,r'}X_rX_{r'}$ describes a ferromagnet and $\beta$ is sufficiently high. At very small $\beta$, the state above is in a weak paramagnetic (WP) phase since there is no long range order parameter for the weak symmetry.

Any strongly symmetric local Lindbladian that has $(1+U)$ as a steady state must have its partner $(1-U)$ also as a steady state. To see this, note that the assumption that $1+U$ is a steady state means  
\begin{equation}
    \mathcal{L}[1+U]=\left(\sum_j[L_j,L_j^\dagger]\right)(1+U)=\left(\sum_j[L_j,L_j^\dagger]\right)+U\left(\sum_j[L_j,L_j^\dagger]\right)=0
\end{equation}
However, $U$ is a global unitary acting everywhere, while the $L_j$ operators are presumed to be local. The above equation cannot hold unless $\sum_j[L_j,L_j^\dagger]=0$. But in this case $\mathcal{L}$ also must stabilize $1-U$. We expect that away from the fixed points, the presence of an SWSSB state in the $+1$ sector also implies an SWSSB state in the $-1$ sector. 

\section{Properties of the marginal fidelity correlator}\label{sproperties}
In this section we provide some derivations for statements in the main text: (1) the implication that nonzero marginal radius-R fidelity correlator implies long range CMI and (2) the error bound for factorizing the two-point marginal fidelity correlator based on decay of mutual information.

\subsection{Proof of long-range CMI}\label{lrcmi}
Here we show that if $O_i$ is charged and $\rho$ is strongly symmetric,
\begin{equation}
    F_i^R(\rho,O_i)\leq 2\sqrt{2}I(A:C|B)^{1/4}
\end{equation}
Therefore, if the state is strongly symmetric and $F_i^R(\rho,Z_i)>0$ uniformly in $R$ until $R\sim\mathcal{O}(\mathrm{poly}\log L)$ then the state has long range CMI, with $I(A:C|B)$ failing to decay exponentially outside of a $\mathcal{O}(\mathrm{poly}\log L)$ region $A$. 

Let $A$ support a charged unitary $O_i$ and $B$ be the buffer, with $R=|A\cup B|$, and let $C=\overline{A\cup B}$. 
By Ref.~\cite{fawzi2015}, there exists a recovery channel $\mathcal{R}_{B\to BC}$ such that $\sigma=\mathcal{R}_{B\to BC}(\rho_{AB})$ obeys
\begin{equation}\label{fawzi}
    \|\rho-\sigma\|_1\leq 2\sqrt{I(A:C|B)}
\end{equation}

Now we evaluate
\begin{equation}
    |F_i(\rho,O_i)-F_i(\sigma,O_i)|=|\mathrm{Tr}(O_i\sqrt{\rho}O_i^\dagger\sqrt{\rho})-\mathrm{Tr}(O_i\sqrt{\sigma}O_i^\dagger\sqrt{\sigma})|
\end{equation}
Adding and subtracting $\mathrm{Tr}(O_i\sqrt{\sigma}O_i^\dagger\sqrt{\rho})$ gives
\begin{align}\label{addsubtract}
    \begin{split}
|F_i(\rho,O_i)-F_i(\sigma,O_i)|&\leq |\mathrm{Tr}(O_i(\sqrt{\rho}-\sqrt{\sigma})O_i^\dagger(\sqrt{\rho})|+|\mathrm{Tr}(O_i\sqrt{\sigma}O_i^\dagger(\sqrt{\rho}-\sqrt{\sigma}))|\\
    &\leq \|\sqrt{\rho}\|_2\|\sqrt{\rho}-\sqrt{\sigma}\|_2+\|\sqrt{\sigma}\|_2\|\sqrt{\rho}-\sqrt{\sigma}\|_2\\
    &\leq 2\|\rho-\sigma\|_1^{1/2}
\end{split}
\end{align}
where the last line follows from $\|\sqrt{\rho}\|_2=\|\sqrt{\sigma}\|_2=1$ and the Powers-Størmer inequality
\begin{equation}
    \|\sqrt{\rho}-\sqrt{\sigma}\|_2^2\leq \|\rho-\sigma\|_1
\end{equation}

Applying this to $\rho=\rho_{ABC}$ and $\sigma=\mathcal{R}_{B\to BC}(\rho_{AB})$ and using $F_i(\rho,O_i)=0$ we get $F_i(\sigma,O_i)\leq 2\|\rho-\sigma\|_1^{1/2}$. Combining this with the Fawzi-Renner recovery estimate (\ref{fawzi}) gives $F_i(\sigma,O_i)\leq 2\sqrt{2}I(A:C|B)^{1/4}$.

The last step is to notice that because $\mathcal{R}_{B\to BC}$ only acts on $B$, it commutes with $O_i$. By monotonicity of fidelity under channels,
\begin{equation}
    F_i(\sigma,O_i)=F_i(R_{B\to BC}(\rho_{AB}),O_i)\geq F_i(\rho_{AB},O_i)=F_i^R(\rho,O_i)
\end{equation}

Using $F_i(\sigma,O_i)\geq F_i^R(\rho,O_i)>0$ gives the desired result.

\subsection{Error bound in $|F_{ij}^R(\rho,O_iO_j^\dagger)-F_i^R(\rho,O_i)F_j^R(\rho,O_j^\dagger)|$}
In this section, we show that 
\begin{equation}\label{decombound}
    |F_{ij}^R(\rho,O_iO_j^\dagger)-F_i^R(\rho,O_i)F_j^R(\rho,O_j^\dagger)|\leq 2(2I(A:C)_{\rho})^{1/4}
\end{equation}
where $I(A:C)$ is the mutual information between regions $A$ and $C$ in $\rho$. This justifies approximating $F_{ij}^R(\rho,O_iO_j^\dagger)$ by the square of the one-point correlator $F_i^R(\rho,O_i)$ when there is exponential decay of MI and translation invariance.

Let $A=R(i),C=R(j)$ and define
\begin{equation}
    \rho_{AC}=\rho_{R(i)\cup R(j)}\qquad \tilde{\rho}_{AC}=\rho_{R(i)}\otimes \rho_{R(j)}
\end{equation}
so $F_{ij}^R(\rho,O_iO_j^\dagger)=\mathrm{Tr}(O_iO_j^\dagger\sqrt{\rho_{AC}} O_i^\dagger O_j\sqrt{\rho_{AC}})$. Then
\begin{equation}
    F_{ij}^R(\rho,O_iO_j^\dagger)-F_i^R(\rho,O_i)F_j^R(\rho,O_j^\dagger)=\mathrm{Tr}(O_iO_j^\dagger\sqrt{\rho_{AC}}O_i^\dagger O_j\sqrt{\rho_{AC}})-\mathrm{Tr}(O_iO_j^\dagger\sqrt{\tilde{\rho}_{AC}}O_i^\dagger O_j\sqrt{\tilde{\rho}_{AC}})
\end{equation}
We add and subtract $\mathrm{Tr}(O_iO_j^\dagger\sqrt{\tilde{\rho}_{AC}}O_i^\dagger O_j\sqrt{\rho_{AC}})$ and use the Powers-Størmer inequality in the same way as in the previous section to get
\begin{equation}
    |F_{ij}^R(\rho,O_iO_j^\dagger)-F_i^R(\rho,O_i)F_j^R(\rho,O_j^\dagger)|
    \leq 2\|\sqrt{\rho_{AC}}-\sqrt{\tilde{\rho}_{AC}}\|_1^{1/2}
\end{equation}

Then using Pinkser's inequality\cite{watrous2018} $I(A:C)_{\rho}=D(\rho_{AC}\|\rho_A\otimes \rho_C)\geq\frac{1}{2}\|\rho_{AC}-\rho_A\otimes\rho_C\|_1^2$ (where $D(\rho_{AC}\|\rho_A\otimes \rho_C)$ is the quantum relative entropy), we obtain (\ref{decombound}).

\section{Evaluation of marginal fidelity for various states}
Here we evaluate the marginal fidelities $F_i^R(\rho,Z_i)$, $F_{ij}^R(\rho,Z_iZ_j)$ for various states, including those relevant to the phases in the main text. In the SP phase, we show that $[F_i^R(\rho,Z_i)]^2$ does not in general well approximate $F_{ij}^R(\rho,Z_iZ_j)$ by explicit calculation. The use of the one-point correlator appears more reliable deep in the SWSSB phase, assuming that the mutual information decays exponentially. We also show that $F_{ij}^R(\rho,Z_iZ_j)$ is lower bounded for symmetry projected Gibbs states, and we obtain a general bound on the error from truncation to $R$ for a broad class of symmetry projected states.

\subsection{Weak (classical) memory}
The state $\rho=\frac{1}{2}(\rho_0+\rho_0^-)$ is a classical memory with a weak $\prod_rZ_r$ symmetry. It is easy to check that $F_{ij}^0(\rho,Z_iZ_j)=F_i^0(\rho,Z_i) = 1$ even though the true two-point and one-point fidelity correlators are exactly zero. However, for any $R>0$, we also get $F_{ij}^R(\rho,Z_iZ_j)=F_i^R(\rho,Z_i)=0$ so there is dramatic improvement going from $R=0$ to $R>0$. 

\subsection{Wandering error cluster}
We first consider the ansatz for the symmetry odd steady state in the 1+1d SP phase: $\rho_{\text{abs}}=\sum_{r,n}c_{r,n}\tilde{Z}_{r,n}\rho_0\tilde{Z}_{r,n}$ where $\tilde{Z}_{r,n}$ is a local charged operator with support $n$ centered at $r$, with $c_r\leq \mathrm{const}\cdot e^{-r/\xi}$ for some $\mathcal{O}(1)$ correlation length $\xi$ and $\sum_{r,n}c_{r,n}=1$. We will assume that the state has a weak translation symmetry, because it is a dressed version of $\rho_W$. Therefore, $c_{n,r}$ only depends on $n$, and we can set $c_{r,n}=a_n/L$ with $\sum_na_n=1$, $a_n\leq \text{const}\cdot e^{-n/\xi}$. 

Let's first consider the simplest case where we have the classical $W$ state $\rho_W=\frac{1}{L}\sum_rZ_r\rho Z_r$, so $a_{n}=\delta_{n,1}$. The one-point fidelity $F_i^R(\rho,\sigma)=0$ due to the strong symmetry, and the true two-point fidelity is $F_{ij}(\rho,\sigma)=\frac{2}{L}$ because there is one contribution from  $X_i=1,X_j=-1\to X_i=-1,X_j=1$ after application of $Z_iZ_j$ and one contribution from $X_i=-1,X_j=1\to X_i=1,X_j=-1$.

For the radius $R$ fidelity, the reduced state has probability $\frac{L-(2R+1)}{L}$ to be in the all $+$ configuration and probability $\frac{1}{L}$ on each of the configurations with a single minus spin. We therefore get
\begin{equation}
    F_i^R(\rho,Z_i)=2\frac{\sqrt{L-(2R+1)}}{L}
\end{equation}
For $R=0$ we recover $F_i^0(\rho,Z_i)\sim \frac{2}{\sqrt{L}}$, which is simply $2\sqrt{p(+)p(-)}$ where $p(\pm )$ is the probability of a spin being $+/-$. 

We also find $F_{ij}^R(\rho,Z_iZ_j)=\frac{2}{L}$ independent of $R$. This is because the contribution from the all $+$ in $R(i)\cup R(j)$ to having an error in $R(i)$ and one in $R(j)$ is zero (since $\rho_W$ consists of only states that have 1 error). The only contribution comes from the probability that an error in $R(i)$ is mapped to an error in $R(j)$ and vice versa. These two terms contribute $\frac{1}{L}$ each. 

In summary, for $\rho_W$, 
\begin{equation}
    F_i^R(\rho,Z_i)=2\frac{\sqrt{L-2R-1}}{L}\to 0 \quad (R\to(L-1)/2)\qquad F_{ij}^R(\rho,Z_iZ_j)=\frac{2}{L}
\end{equation}

Indicating that $[F_i^R(\rho,Z_i)]^2$ may overestimate  $F_{ij}^R(\rho,Z_iZ_j)$ for these kinds of states. More generally, let $p_{i/j}^R$ be the total probability that the excitation overlaps $R(i/j)$:
\begin{equation}
    p_i^R=\sum_{r,n:\text{ supp}(\tilde{Z}_{r,n})\cap R(i)\neq\emptyset}c_{r,n}\qquad p_j^R=\sum_{r,n:\text{ supp}(\tilde{Z}_{r,n})\cap R(j)\neq\emptyset}c_{r,n}
\end{equation}
Assuming that $2R+n\ll L$ so that wraparound effects are negligible, we have 
\begin{equation}
    p_i^R\sim \frac{1}{L}\sum_na_n(2R+n)=\frac{1}{L}(2R+\langle n\rangle)=p_j^R\qquad \langle n\rangle =\sum_na_nn
\end{equation}
Therefore, the probability of seeing all $+$ in region $R(i)$ is $1-p_i^R$ and we get
\begin{equation}
    F_i^R(\rho,Z_i)\sim 2\sqrt{\left(1-\frac{2R+\langle n\rangle }{L}\right)\frac{a_1}{L}}+\mathcal{O}(1/L)
\end{equation}
so we get a result similar to the previous case, with $2R+1\to 2R+\langle n\rangle$. The two-point correlator remains $\frac{2a_1}{L}$ up to errors exponentially small in $|i-j|$, because it again only picks up processes where an excitation hops between $i$ and $j$ due to the application of $Z_iZ_j$.

\subsection{Simple classical Gibbs state}
Here we work out exactly the one-point fidelity correlator for the state
\begin{equation}
    \rho=e^{-\beta H}/Z
\end{equation}
with $H=-\sum_{r,r'}X_rX_{r'}$ a nearest neighbor Ising ferromagnet in the $X$ basis. We will show that $F_i^R(\rho,Z_i)$ converges exactly for $R>1$ and $F_{ij}^R(\rho,Z_iZ_j)\sim [F_i^R(\rho,Z_i)]^2$ everywhere away from $\beta_c$ (even for $\beta>\beta_c$). 

In this case, 
\begin{equation}
    F_i^R(\rho,Z_i)=\frac{1}{Z}\mathrm{Tr}(Z_ie^{-\beta H/2}Z_ie^{-\beta H/2})=\mathrm{Tr}(e^{-\beta O}e^{-\beta H})=\frac{1}{Z}\langle e^{-\beta O}\rangle_{\rho}
\end{equation}
where $O=\sum_{r'}X_iX_{r'}$ is a sum of Ising couplings between the site $i$ and its neighbors. Note that $O$ is symmetric under the weak $\mathbb{Z}_2$ $\prod_rZ_r$ symmetry, so its connected correlation functions cluster away from $\beta_c$. This is true for any $H$ that commutes with the weak $\mathbb{Z}_2$ symmetry. Also note that we get the exact same result for any choice of $R\geq 1$, so the marginal fidelity converges immediately. This generalizes to any finite range Hamiltonian: we just need $R\geq \ell$ where $\ell$ is the range of the Hamiltonian. Quasilocal $H$ would give error exponentially small in $R$. 

Adding the global parity projector $(1+U)e^{-\beta H}/Z$ minimally changes the result, because when we take a reduced density matrix on $R(i)$, the contribution from $Ue^{-\beta H}/Z$ is exponentially small in volume (see the discussion in Sec.~\ref{bounderror})

\subsection{(Symmetry projected) Gibbs states}
These states make up a standard class of SWSSB states. We will first show that there is a lower bound on the two-point fidelity correlator when the inverse temperature $\beta$ is finite, and then we will show that the marginal radius $R$ fidelity correlator approximates the true fidelity correlator with error exponentially small in $R$ under certain local indistinguishability assumptions.
\subsubsection{Lower bound on fidelity correlator}
We first consider $\tau=\frac{1}{Z}e^{-\beta H}$ and we will then add the symmetry sector projector. We will first show that the true one-point fidelity (and therefore also the marginal fidelity) is lower bounded by an $\mathcal{O}(1)$ constant as long as $\beta$ is finite and $H$ is local. The form of the lower bound is similar to the exact result in the classical case computed above.

The full one-point fidelity is
\begin{equation}
    F_i(\tau,Z_i)=\mathrm{Tr}\left(\sqrt{\tau}\sqrt{\tau'}\right)=e^{-D_{1/2}(\tau\|\tau')/2}
\end{equation}
where $\tau'=\frac{1}{Z}e^{-\beta Z_iH Z_i}$ and $D_{\alpha}(\tau\|\tau')$ is the Petz Rényi divergence\cite{petz1986} $D_{\alpha}(\rho\|\sigma)=\frac{1}{\alpha-1}\log \mathrm{Tr}\left(\rho^\alpha\sigma^{(1-\alpha)}\right)$ with $\alpha=1/2$. Note that $\lim_{\alpha\to 1}D_{\alpha}(\rho\|\sigma)$ recovers the quantum relative entropy $D(\rho\|\sigma)$\cite{muller2013}. We use the fact that $D_{\alpha}(\rho\|\sigma)$ is nondecreasing as a function of $\alpha$ for fixed $\rho,\sigma$ (see proof below) to obtain 
\begin{equation}
    F_i(\tau,Z_i)\geq e^{-D(\tau\|\tau')/2}
\end{equation}
Let us denote $Z_iHZ_i=H+V_i$ where $V_i$ is an operator supported near $i$. Since $\log\rho=-\beta H-\log Z$ and $\log \rho'=-\beta (H+V)-\log Z$ (since $Z_i$ is unitary), we get
\begin{equation}
    D(\tau\|\tau')=\beta\mathrm{Tr}(\tau((H+V)-H))=\beta\langle V_i\rangle_\tau\leq \beta\|V_i\|
\end{equation}
It follows that
\begin{equation}
F_i(\tau,Z_i)\geq e^{-\beta \|V_i\|/2}
\end{equation}
so $F_i(\tau,Z_i)$ is lower bounded by an $\mathcal{O}(1)$ constant by the bound on $\beta\|V_i\|/2$.

All that remains is to prove that for fixed $\rho$ and $\sigma$ and $\beta>\alpha$, $D_{\beta}(\rho,\sigma)\geq D_{\alpha}(\rho,\sigma)$. To do so, we will relate $D_{\alpha}(\rho,\sigma)$ to a classical Renyi divergence of probability distributions, and then we can directly apply Theorem 3 of \cite{vanerven2014}.

First diagonalize $\rho$ and $\sigma$ as $\rho=\sum_i\lambda_i|\psi_i\rangle\langle\psi_i|$ and $\sigma=\sum_i\mu_j|\phi_j\rangle\langle\phi_j|$. Then 
\begin{equation}
    \mathrm{Tr}\left(\rho^\alpha \sigma^{(1-\alpha)}\right)=\sum_{ij}\lambda_i^{\alpha}\mu_j^{(1-\alpha)}|\langle\psi_i|\phi_j\rangle|^2=\sum_{ij}p_{ij}^\alpha q_{ij}^{(1-\alpha)}
\end{equation}
where 
\begin{equation}
    p_{ij}=\lambda_i|\langle \psi_i|\phi_j\rangle|^2\qquad q_{ij}=\mu_j|\langle\psi_i|\phi_j\rangle|^2
\end{equation}
are probability distributions: $\sum_{ij}p_{ij}=\sum_{ij}q_{ij}=1$ and $p_{ij},q_{ij}\in \mathbb{R}^+$. Now we can directly apply Theorem 3 of \cite{vanerven2014}.

For the fidelity correlator defined by the Uhlmann fidelity rather than the Holevo fidelity, a similar argument follows from using the ``new" Renyi divergence\cite{muller2013}
\begin{equation}
    \tilde{D}_{\alpha}(\rho\|\sigma)=\frac{1}{\alpha-1}\log\mathrm{Tr}\left[\left(\sigma^{(1-\alpha)/2\alpha}\rho\sigma^{(1-\alpha)/2\alpha}\right)^{\alpha}\right]
\end{equation}
which also converges to $D(\rho\|\sigma)$ and $\alpha\to\ 1$. Since the Holevo fidelity is lower bounded by the Uhlmann fidelity, the above bound using Petz divergence is slightly stronger. 

Now we turn to the projected Gibbs state $\propto (1+U)e^{-\beta H}=\lim_{\gamma\to\infty}e^{-\beta H+\gamma U}/Z'$ by defining $\tilde{H}=H-(\gamma/\beta)U$. Since $Z_i$ is symmetry odd, we have $Z_i\tilde{H}Z_i=H+V_i+(\gamma/\beta)U$, so
\begin{equation}
    D(\rho\|\rho')=\beta\mathrm{Tr}(\rho(V_i+2(\gamma/\beta)U))\to\infty\qquad \gamma/\beta\to\infty
\end{equation}

So $F_i(\rho,Z_i)$ is lower bounded by zero. On the other hand for a symmetric operator like $Z_iZ_j$, we have $Z_iZ_j\tilde{H}Z_iZ_j=H+V_i+V_j-(\gamma/\beta)U$, giving 
\begin{equation}
    F_{ij}(\rho,Z_iZ_j)\geq e^{-\beta\|V_i+V_j\|/2}
\end{equation}

By monotonicity of fidelity, the marginal radius $R$ fidelity is also lower bounded.

\subsubsection{Bounding the error of the marginal fidelity}\label{bounderror}
For states with exponential decay of CMI, including all classical Gibbs states, we can also bound the error between the true fidelity correlator and the marginal radius $R$ fidelity as long as we make an additional assumption of local indistinguishability. We define
\begin{equation}
    \rho=P_+\tau P_++P_-\tau P_-\qquad p_{\pm}=\mathrm{Tr}(P_{\pm}\rho)\qquad\rho_{\pm}=\frac{P_{\pm}\rho P_{\pm}}{p_{\pm}}
\end{equation}
where $P_{\pm}=\frac{1\pm U}{2}$. Then our assumption of local insdistinguishability means that
\begin{equation}\label{ind}
    \|\rho_{+,AB}-\rho_{-,AB}\|_1\leq \eta_R\qquad |F(\rho_+,O)-F(\rho_-,O)|\leq \zeta
\end{equation}
where $\rho_{\pm,AB}=\mathrm{Tr}_C(\rho_{\pm})$ with the same geometry as before: $AB=R(i)\cup R(j)$ and $C=\overline{AB}$. We assume that $\eta_R$ is exponentially decaying in $R$ and $\zeta$ is exponentially small in system size for $O$ that is symmetry even. These assumptions are most natural when $\rho$ is a Gibbs state with weak symmetry $U$.

From the definitions above, we have $\rho=p_+\rho_++p_-\rho_-$. Let $O$ be an even unitary operator i.e. $O=Z_iZ_j$. Let
\begin{equation}
    F_{ij}(\rho,O)=\mathrm{Tr}(O\sqrt{\rho}O^\dagger\sqrt{\rho})\qquad F_{ij}^R(\rho,O)=F_{ij}(\rho_{AB},O)
\end{equation}
We assume that $\rho$ has exponential decay of CMI (note that $\rho_{\pm}$ do not have exponential decay of CMI), giving
\begin{equation}
    0\leq F_{ij}^R(\rho,O)-F(\rho,O)\leq 2\sqrt{\delta_R}
\end{equation}
where $\delta_R\leq 2\sqrt{\epsilon_R}$ and $\epsilon_R$ describes the exponential decay of CMI (see Sec.~\ref{lrcmi}). We can then use this to prove a similar result for the projected state $\rho_+$.

Because $\rho$ is block diagonal, $\sqrt{\rho}$ is also block diagonal:
\begin{equation}
    \sqrt{\rho}=\sqrt{p_+}\sqrt{\rho_+}+\sqrt{p_-}\sqrt{\rho_-}
\end{equation}
Since $[O,U]=0$, $O$ cannot mix the blocks. Therefore, the fidelity decomposes as 
\begin{equation}
    F_{ij}(\rho,O)=p_+F_{ij}(\rho_+,O)+p_-F_{ij}(\rho_-,O)
\end{equation}

The marginal state also decomposes as
\begin{equation}
    \rho_{AB}=p_+\rho_{+,AB}+p_-\rho_{-,AB}
\end{equation}
But after tracing out $C$, $\rho_{+,AB}$ and $\rho_{-,AB}$ may no longer be orthogonal blocks. So $F_{ij}^R(\rho,O)$ does not necessarily decompose as $F_{ij}(\rho,O)$ did. This is where we must invoke local indistinguishability (\ref{ind}). The assumption of local indistinguishability together with the same manipulations and Powers-Størmer inequality as in Sec.~\ref{lrcmi} gives,
\begin{equation}
    |F_{ij}^R(\rho_+,O)-F_{ij}^R(\rho_-,O)|\leq 2\sqrt{\eta_R}
\end{equation}

Since $F_{ij}(\rho,O)$ is the sector weighted average of $F_{ij}(\rho_{\pm},O)$,
\begin{equation}
    |F(\rho_+,O)-F(\rho,O)|\leq p_-\zeta\leq \zeta
\end{equation}

Similarly, $\rho_{AB}$ is locally close to either sector marginal:
\begin{equation}
    \|\rho_{AB}-\rho_{+,AB}\|_1=p_-\|\rho_{-,AB}-\rho_{+,AB}\|\leq \eta_R
\end{equation}
Again applying the manipulations in Sec.~\ref{lrcmi} gives
\begin{equation}
    |F_{ij}^R(\rho,O)-F_{ij}^R(\rho_+,O)|\leq 2\sqrt{\eta_R}
\end{equation}

Putting everything together,
\begin{equation}
    F_{ij}^R(\rho_+,O)-F_{ij}(\rho_+,O)=(F_{ij}^R(\rho_+,O)-F_{ij}^R(\rho,O))+(F_{ij}^R(\rho,O)-F_{ij}(\rho,O))+(F_{ij}(\rho,O)-F_{ij}(\rho_+,O))
\end{equation}
It follows that
\begin{equation}
    |F_{ij}^R(\rho_+,O)-F_{ij}(\rho_+,O)|\leq 2\sqrt{\delta_R}+2\sqrt{\eta_R}+\zeta
\end{equation}
which are all exponentially decaying in $R$. 

\subsection{Partially dead state}
Finally, we will study $\rho=s\rho_-+(1-s)\rho_0$ where $s$ can be interpreted as a survival probability. These kinds of states may be seen in the even symmetry sector of the strong $\mathbb{Z}_2$ symmetry in the models in the main text. We will calculate the marginal fidelities for the 1+1d case but it is straightforward to generalize to 2+1d. 

Recall that $\rho_-\propto \frac{1+U}{2}-\rho_0$ and $\rho_0=|+\rangle\langle +|$. Therefore, the state is diagonal in the $X$ basis with $p_0=1-s$ for the state $|+\rangle\langle +|$ and $p_x=\frac{s}{2^L}$ for all other even parity configurations $s$. The true fidelity is given by
\begin{equation}
    F_{ij}(\rho,\sigma)=2\sqrt{\frac{s(1-s)}{2^L-1}}+\frac{s}{2^L-1}(2^L-2)\sim s+\mathcal{O}(2^{-L})
\end{equation}

We now evaluate the marginal fidelity correlator with radius $R$. Assuming that $|i-j|>2R$, the reduced state $\rho_{R(i)\cup R(j)}$ has the following $X$ basis probabilities:
\begin{equation}
    p_{0,R}=1-s+s\frac{2^{L-2(2R+1)-1}-1}{2^{L}-1}\qquad p_{x\neq 0,R}=s\frac{2^{L-2(2R+1)-1}}{2^L-1}
\end{equation}
This is because after tracing out $\overline{R(i)\cup R(j)}$, almost every local configuration has the same number of compatible global even-parity completions except the $0$ configuration. We then have
\begin{equation}
    F_{ij}^R(\rho,\sigma)=2\sqrt{p_{0,R}p_{x\neq 0,R}}+(2^{2(2R+1)}-1)p_{x\neq 0,R}
\end{equation}
In the limit $N\gg2(2R+1)$, we get $p_{0,R}\to 1-s+s2^{-2(2R+1)}$ and $p_{x\neq 0,R}\to s2^{-2(2R+1)}$ so
\begin{equation}
    F_{ij}^R(\rho,\sigma)=s+2\sqrt{s(1-s)}2^{-(2R+1)}-2s2^{-2(2R+1)}+\mathcal{O}(2^{-3(2R+1)})
\end{equation}
which approaches $s$ exponentially in $R$. However, for $R$ finite, at very small $s$ the leading contribution to $F_{ij}^R(\rho,\sigma)$ is $2\sqrt{s}2^{-(2R+1)}$. The leading contribution becomes $s$ for $s>\frac{1}{1+2^{4R}}$. $\rho$ may still be in the SWSSB phase if $s\sim 1-\frac{1}{\mathrm{poly}(L)}$. In this case we would need $R\sim\mathcal{O}(\mathrm{poly}\log L)$ to have $|F_{ij}^R(\rho,Z_iZ_j)-F_{ij}(\rho,Z_iZ_j)|\to 0$ as $L\to \infty$.

\section{Pair flip Toom dynamics}
In this section we specify the precise classical $X$-basis update rules for the pair-flip Toom dynamics.
\begin{enumerate}
    \item Pick a site at random. 
    \item If it is not anti-aligned with both its north and east neighbors, do nothing. 
    \item If the site is anti-aligned with its north and east neighbors, it is ``Toom unstable." Compute the change in domain wall length from flipping the site together with each of its eight neighbors (including diagonal ones). This can be computed locally by looking at the 24 bonds around the target site and its eight neighbors. 
    \item Collect the subset of these pair flips $PF_{\text{strict}}$ that lower the total domain wall length and also the subset $PF_{\text{chill}}$ that do not increase the domain wall length. $PF_{\text{strict}}\subset PF_{\text{chill}}$.
    \item If the relevant subset of flips ($PF_{\text{strict}}$ or $PF_{\text{chill}}$) is empty, do nothing.
    \item Otherwise, choose uniformly among the elements of $PF_{\text{strict}}$ for strict pair-flip Toom or $PF_{\text{chill}}$ for chill pair-flip Toom.
\end{enumerate}

\begin{figure}[t]
   \centering
   \includegraphics[width=.7\columnwidth]
   {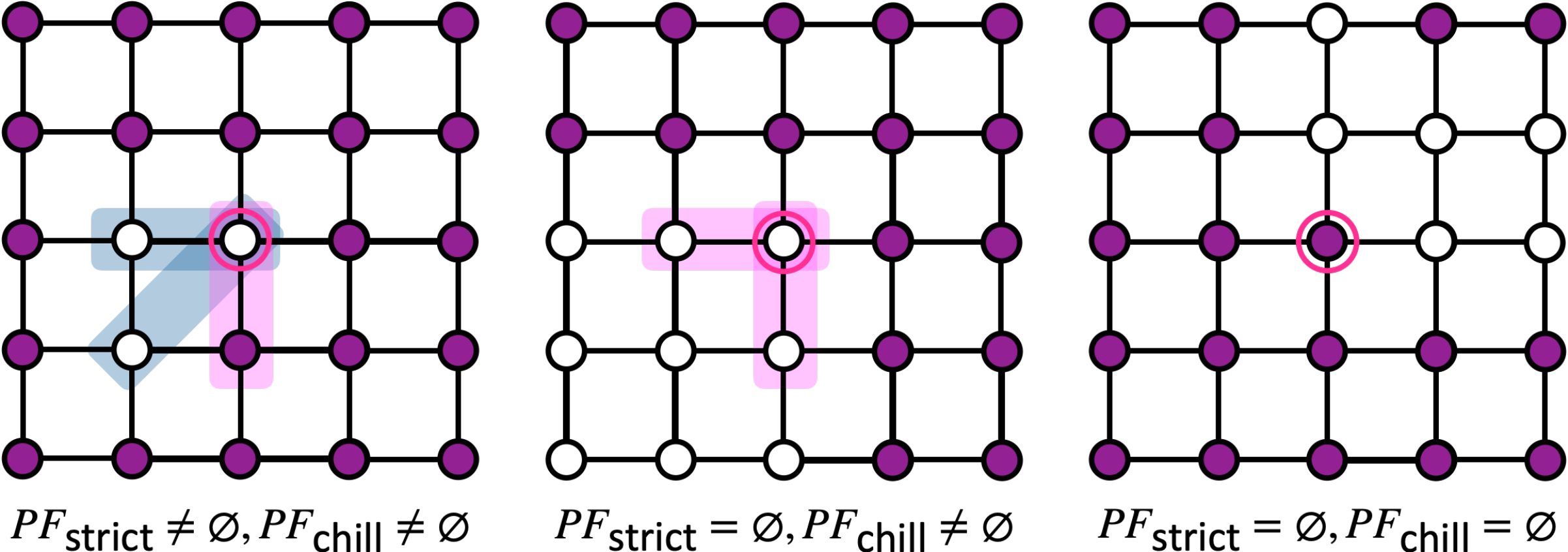}
   \caption{The purple circles represent $+$ spins and the white circles represent $-$ spins. We focus on the Toom unstable spin circled in pink, and show the available strict pair-flip Toom moves in blue and chill pair-flip Toom rules in pink.} 
   \label{fig:example_toom}
\end{figure}

It is straightforward to find configurations for which $PF_{\text{strict}}=\emptyset$: any isolated single minority spin or corner of a minority square with side length $\geq 3$ is frozen. We also show below a configuration where $PF_{\text{chill}}=\emptyset$ in Fig.~\ref{fig:example_toom}.

\begin{figure}[t]
   \centering
   \includegraphics[width=.7\columnwidth]
   {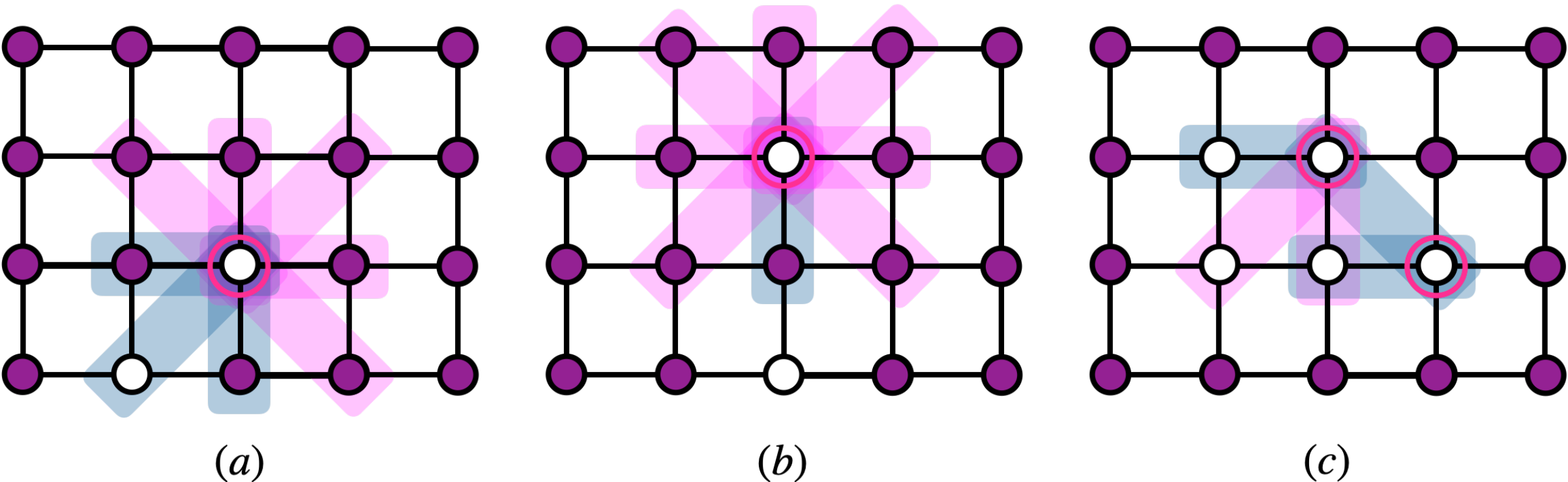}
   \caption{We take the same coloring conventions as in Fig.~\ref{fig:example_toom} and focus on the Toom unstable spin circled in pink, assuming that beyond the illustrated ones are all $+$. $(a)$ and $(b)$ give examples of configurations where strict pair-flip Toom pushes the two minority spins together into a larger domain or completely annihilates the two minority spins, while chill pair flip Toom allows the minority spins to get farther apart with probability $5/8$ ($a$) or $7/8$ ($b$). In $(c)$, chill pair-flip Toom can split the minority domain into two small islands, that can then wander away from each other.} 
   \label{fig:strictchill}
\end{figure}

As mentioned in the main text, one possible explanation for why strict pair-flip Toom appears to have a stable SP phase and chill pair-flip Toom may not is that chill pair-flip Toom allows for dangerous branching processes where small islands break off of their main domain and then can easily diffuse around the system and continue to branch. We illustrate this in Fig.~\ref{fig:strictchill}. 

\section{Steady states for 1+1d and 2+1d SWSSB dynamics}
Here we consider the fixed point $\alpha=1$ SWSSB dynamics in 1+1d with the triplet update rule and 2+1d with the plaquette update rule and show that they are precisely 
\begin{align}
\begin{split}
    &\mathrm{span}\left(\rho_0,\frac{1+U}{2}-\rho_0,\frac{1-U}{2}\right)\qquad (1+1d)\\
    &\mathrm{span}\left(\rho_0,\rho_0^-,\frac{1+U}{2}-\rho_0-\rho_0^-,\frac{1-U}{2}\right)\qquad (2+1d)
\end{split}
\end{align}

First we will check that in 1+1d, there are only three connected classes consisting of $\rho_0$, $\{\text{even parity states}\}\setminus \rho_0$ and $\{\text{odd parity states}\}$. The $\rho_0$ class is clearly isolated because there are no transitions from two minus states to $\rho_0$ and no transitions out of $\rho_0$. To show that the other classes are connected, note that 
\begin{enumerate}
    \item Single $-$ spins can move.
    \item Three $-$ spins together can be reduced to one: $(---)\leftrightarrow(-++),(+-+),(++-)$
    \item Pairs of $-$ spins can move because we scramble $(--+),(-+-),(+--)$.
\end{enumerate}
For every odd parity state, we can choose one $-$ spin to be the seed and move other $-$ spins to the seed to form a local $(---)$ and then use (2) to remove two $-$ spins. Repeating this procedure reduces any odd configuration to a single $-$ spin, and since it can move anywhere by (1), the whole odd sector is connected.

For every even parity configuration with $\geq 2$ minus spins, we can similarly reduce $-$ spins pairwise until two remain. Then using (3) we can connect all states with two $-$ spins, so all even configurations besides $\rho_0$ are also connected. 

In the 2+1d case, we have four classes: $\rho_0$, $\rho_0^-$, $\frac{1+U}{2}-\rho_0-\rho_0^-$, and $\frac{1-U}{2}$. We use the following gadgets:
\begin{enumerate}
    \item Single $-$ spins can move within a plaquette
    \item A pair of $-$ spins can be rearranged inside a plaquette 
    \item One $-$ spin can transition to three minus spins in a plaquette and vice versa
\end{enumerate}

It is clear that $\rho_0$ and $\rho_0^-$ are completely decoupled from the rest of the states and form their own class. In the odd sector, any configuration can be reduced to a single $-$ spin, which can then be moved around by (1). Therefore, the odd sector is connected. 

In the even sector, outside of $\rho_0$ and $\rho_0^-$, every state contains both $+$ spins and $-$ spins. We can always bring all of the $-$ spins together by (1) and (2), and then annihilate them pairwise with (3) until we have two remaining $-$ spins. These can then move throughout the system, so the minus sector $\setminus\{\rho_0,\rho_0^-\}$ is connected. 

All that remains is to show that for any $S,S'$ in the same connected class, the rate of $S\to S'$ is the same as the rate of $S'\to S$. This follows immediately from the observation that all of the transitions above, both in 1+1d and 2+1d, have equal probability in the forward direction and backward direction (i.e. the rate of $(---)\to (-++)$ is the same as the rate of $(-++)\to (---)$ in the 1+1d case since odd parity state scramble uniformly between each other). Therefore, the steady states are the uniform distributions in each connected class.

\section{Additional numerical results for the 2+1d model}

\begin{figure}[t]
   \centering
   \includegraphics[width=1.\columnwidth]
   {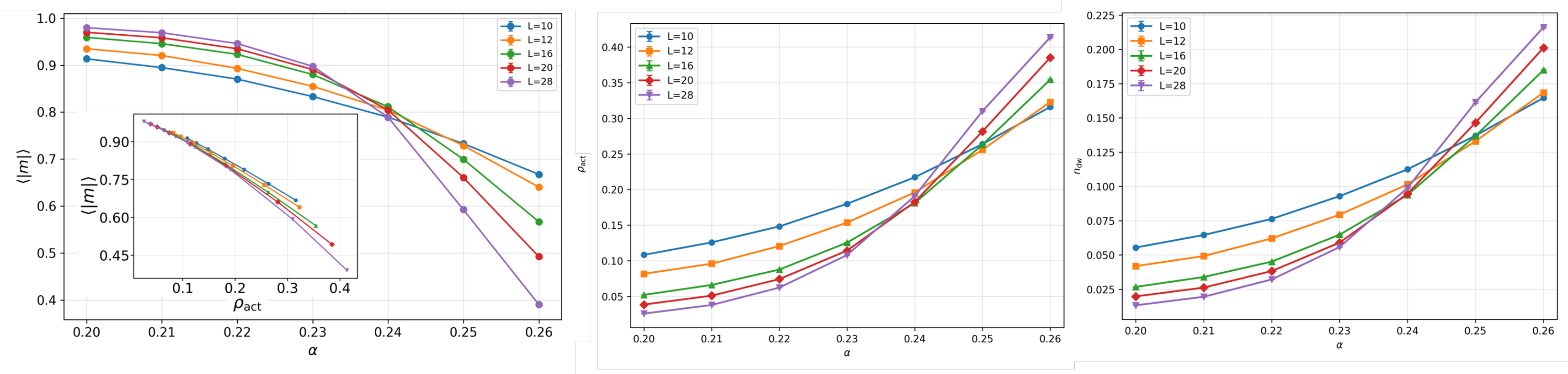}
   \caption{$\langle |m|\rangle$, $\rho_{\mathrm{act}}$, and $n_{\mathrm{dw}}$ as a function of $\alpha$, averaged over 500 runs and 200 late time samples per run. The first quantity is more directly a probe of the WTSSB-WP  transition while the other two are more directly probes of the SP-SWSSB transition. In the inset, we plot $\langle|m|\rangle$ against $\rho_{\mathrm{act}}$ to probe how anticorrelated they are.} 
   \label{fig:otherdata}
\end{figure}

Here we present additional numerical results for the 2+1d strict Toom phase transition. Our main results are in Fig.~\ref{fig:otherdata}. The density of minus spins, which was used in the 1+1d case as an order parameter, is not a good order parameter in this case because the weak $\mathbb{Z}_2$ symmetry forces $\sum_r\langle X_r\rangle =0$. Instead, we measure the following quantities:
\begin{itemize}
    \item $\langle |m|\rangle:$ absolute value of magnetization, where $m=\sum_rX_r$.
    \item $\rho_{\mathrm{act}}$: active density, which is the density of plaquettes with configuration not $(++++)$ or $(----)$
    \item $n_{\mathrm{dw}}$: domain wall density  $\frac{1}{2L^2}\sum_{r,r'}\frac{1-X_rX_{r'}}{2}$
\end{itemize}

As clear from Fig.~\ref{fig:otherdata}, all of these quantities behave similarly. $\langle |m|\rangle$ fits nicely to 2d Ising critical exponents $\beta=1/8,\nu=1$ in other examples of WT-WS transitions such as 2+1d pair flip Glauber dynamics for $e^{\beta\sum_{r,r'}X_rX_{r'}}/Z$. If there is a direct transition between SP/WTSSB and SWSSB/WP then $\langle |m|\rangle$ may behave a bit differently because it gets pinned at 1 in the SP phase; in a typical WTSSB phase, $\langle |m|\rangle$ may be $\mathcal{O}(1)$ but not quite $1$. On the other hand, if there is an intervening SWSSB/WTSSB phase, then we would expect $\langle |m|\rangle$ to follow Ising critical exponents. All three quantities for these system sizes give rather poor scaling collapses using Ising critical exponents. More numerical work with larger system sizes is needed to obtain a convincing set of critical exponents. Recall that the $m_{\infty}$ plots in the main text indicate a transition at $\alpha_c>0.15$, and the Binder plot indicates that the WTSSB to WP transition occurs at $\alpha_m\sim 0.2428$.

In the inset, we also plotted $\langle |m|\rangle$ against $\rho_{\mathrm{act}}$ to probe how anticorrelated they are. For $\langle|m|\rangle>0.8, \rho_{\mathrm{act}}<0.175$, the system sizes $L=16,20,28$ appear to lie on top of each other. This is also where these larger system sizes approximately cross each other in the $\langle |m|\rangle$ vs $\alpha$ and $\rho_{\mathrm{act}}$ vs $\alpha$ plots.

\section{Estimate of $\alpha_c$ in the 1+1d phase transition}
In this section, we estimate $\alpha_c$ in the 1+1d process based on the balance between ``escape" probability and ``collapse" probability. A single error cannot be annihilated, and drifts to the left under the SP dynamics. The SWSSB dynamics allows for a single error to branch into three neighbors, giving the configuration $S=(++---++)$. We will label the sites by $r\in[0,6]$. The dangerous processes are processes where errors escape from the main bubble and can grow into their own error bubbles (contributing to $\Gamma_{\text{escape}}$. On the other hand, updates can also collapse this size three bubble into a single error (contributing to $\Gamma_{\text{collapse}}$). 

Let's consider first the absorbing dynamics, which occur at a rate $1-\alpha$ and flip sites $r-1,r$ when $X_r=-1$. We get:
\begin{equation}
    r=2: S\to (+-+--++)\qquad r=3: S\to (++++-++)\qquad r=4:S\to (++-++++)
\end{equation}
so the absorbing rule gives $\Gamma^\text{abs}_{\text{escape}}\sim 1-\alpha, \Gamma^{\text{abs}}_{\text{collapse}}\sim 2(1-\alpha)$. 

The SWSSB processes act on the five triples overlapping with the error bubble: $(++-)$, $(+--)$, $(---)$, $(--+)$, $(-++)$. The central $(---)$ triple scrambles with the one-minus states, so with probability $3/4$, it collapses the size three bubble into a single error. This gives $\Gamma^{\text{SWSSB}}_{\text{collapse}}\sim\frac{3}{4}\alpha$. The SWSSB rule acting on $(++-)$ and $(-++)$ are more dangerous. For example, 
\begin{equation}
    (++-)\to \{(++-),(+-+),(-++),(---)\}
\end{equation}
and only the $(++-)$ outcome is harmless. The other three outcomes allow for either escape or expansion, so we get a contribution of $2\frac{3}{4}\alpha$ to $\Gamma^{\text{SWSSB}}_{\text{escape}}$. For the $(+--)$ and $(--+)$, there is a $2/3$ chance for each to have an escaping error, so they contribute $2(2/3)\alpha$ to $\Gamma^{\text{SWSSB}}_{\text{escape}}$. Putting together $\Gamma_{\text{escape}}=\Gamma^{\text{abs}}_{\text{escape}}+\Gamma^{\text{SWSSB}}_{\text{escape}}$ and $\Gamma_{\text{collapse}}=\Gamma^{\text{abs}}_{\text{collapse}}+\Gamma^{\text{SWSSB}}_{\text{collapse}}$, we get 
\begin{equation}
    \Gamma_{\text{escape}}-\Gamma_{\text{collapse}}=(1-\alpha)-2(1-\alpha)-\frac{3}{4}\alpha+2\frac{3}{4}\alpha+2\frac{2}{3}\alpha\to \alpha_c=\frac{12}{37}\sim0.32
\end{equation}
where $\alpha_c$ is the value of $\alpha$ for which $\Gamma_{\text{escape}}-\Gamma_{\text{collapse}}=0$. In the above calculation, we weighed breaking off an error separated from the main bubble by one spin vs two spins to be the same. A more careful analysis would weigh breaking off an error to a distance two from the main bubble more heavily than breaking off an error to a distance one from the main bubble. For example, dividing the contribution above to $\Gamma_{\text{escape}}$ from errors one site away from the main bubble in half (because they can easily get pulled back into the main bubble) would give $\alpha_c=9/16\sim 0.56$. The actual numerically obtained $\alpha_c$ is $\alpha_c\sim 0.44$, which lies in between these two rough estimates.

\bibliography{lindbladian}